\documentclass[aps,twocolumn,floats,prb]{revtex4}
\usepackage{graphics,dcolumn}
\begin{document}

%======================================================================
\title{A first principles simulation of rigid water}
%======================================================================
\author{Markus Allesch}
\affiliation{Department of Theoretical Physics,
Graz University of Technology, Petersgasse 16, A-8010 Graz, Austria}
\affiliation{Lawrence Livermore National Laboratory,
P.O. Box 808, Livermore, CA 84559}
\author{Eric Schwegler}
\author{Fran\c{c}ois Gygi}
\author{Giulia Galli}
\affiliation{Lawrence Livermore National Laboratory, 
P.O. Box 808, Livermore, CA 84559}

\begin{abstract}
We present the results of Car-Parrinello (CP) simulations 
of water at ambient conditions and under pressure, using 
a rigid molecule approximation. Throughout our 
calculations, water molecules were maintained at a fixed
intramolecular geometry corresponding to the average 
structure obtained in fully unconstrained simulations. 
This allows us to use larger time steps than those adopted 
in ordinary CP simulations of water, and thus to access 
longer time scales. In the absence of chemical reactions 
or dissociation effects, these calculations open the way to 
{\it ab initio} simulations of aqueous solutions that require
timescales substantially longer than presently feasible
(e.g.~simulations of hydrophobic solvation). Our 
results show that structural properties and diffusion 
coefficients obtained with a rigid model are in better 
agreement with experiment than those determined with fully 
flexible simulations. Possible reasons responsible for this 
improved agreement are discussed.
\end{abstract}

\date{\today}
\maketitle

\section{Introduction}

The importance of water in many areas of science has 
motivated a large number of experimental and theoretical 
investigations of the liquid. 
However, it is only recently that x-ray and neutron 
diffraction measurements have come to an overall agreement 
for properties such as the structure of water at ambient conditions
\cite{jsorenson00,asoper00b}.
In addition, several details of the water structure remain 
the subject of debate, and many dynamical properties of the 
liquid are not yet well understood.

Theoretical models have played an important role in the 
interpretation of experimental measurements and in understanding 
the physical properties of water \cite{bguillot02}. Over the last thirty 
years, rather accurate empirical force fields 
have been developed, 
which can reproduce not only the structure but also many dynamical 
properties of the liquid
\cite{fstillinger74,hberendsen81,mmahoney00}. 
Although empirical models work well for pure water under ambient 
conditions, they are usually difficult to generalize to
complex solutions or thermodynamic states
far from ambient conditions. For example, the majority of
empirical water models that are in current use employ 
potentials that do not change depending on the environment. 

In recent years, it has become possible to simulate the 
properties of a liquid entirely from first principles, without 
having to resort to fitted potentials. This is due in large 
part to the development of the Car-Parrinello (CP) 
method \cite{rcar85} along with the continual increase in high 
performance computing resources.  Although rather 
accurate and with the potential of being a truly predictive tool, CP 
simulations are much more computationally intensive than classical 
simulations.

The case of water is particularly demanding for CP simulations.
The ionic vibrational spectrum of the liquid exhibits high 
frequency modes, i.e. O-H stretch (3200 to 3600 cm$^{-1}$) and 
H-O-H bending modes ($\sim$1600 cm$^{-1}$). Therefore, in order 
to avoid a coupling between the ionic and electronic degrees 
of freedom, which could cause severe inaccuracies in a 
CP simulation, a relatively small fictitious electronic mass 
( $\mu\simeq$ 400 a.u. for protonated water) needs to be used
\cite{jgrossman03}.
In turn, the use of small values of $\mu$, together with the 
high kinetic energy cutoff required to describe the oxygen 
pseudopotential in a plane wave description, necessitates the use 
of small integration time steps. The time step may need to be 
as small as 0.08 fs, which is approximately ten times smaller 
than what is often used in classical MD simulations with empirical 
inter-atomic potentials.  
This poses a severe restriction on the time scales that can be 
accessed in CP simulations of water.

We note that when using Born-Oppenheimer (BO) dynamics 
(where the total energy of the system is minimized at each 
ionic step), one can use larger time steps than in CP 
simulations, since electronic degrees of freedom are not 
propagated at the same time as ionic 
coordinates. However, the accuracy required to reduce systematic 
errors on the ionic forces so as to have conservative dynamics 
is such that large number of iterations are usually 
necessary to minimize the Kohn-Sham energy at each ionic step. 
Therefore, the gain in efficiency obtained with a large time step 
is more than counter-balanced by the computational time requirement for 
total energy minimizations.

It is interesting to note that the problem of integrating fast
vibrational modes in simulations of liquid water has also been 
encountered in classical MD simulations, where the most common 
approach has been to completely eliminate the high-frequency 
intra-molecular motion by using bond length and angle constraints
\cite{wgunsteren82,tschlick97}.
Based on this approach, 
a variety of classical water potentials, e.g. the TIP 
series \cite{mmahoney00}, are capable of accurately 
reproducing many of 
the interesting properties of water. In particular, results 
obtained with the rigid water TIP5P potential are in very good 
agreement with a variety of experimental measurements 
such as the 
structure, the temperature of maximum density, 
diffusion, as well as dielectric properties 
\cite{mmahoney00,mmahoney01,mmahoney01b}. 

With the aim of investigating how to increase the integration
time step in CP simulations of water and thus access larger time 
scales, we have carried out calculations using a rigid water 
approximation. 
In this paper, we present the results of these
simulations and compare them 
to those obtained with flexible 
water molecules (i.e. without imposing any constraints on the 
geometry of the molecules in the liquid), and we discuss the 
effect of a rigid model on the structural properties of water 
at ambient and high pressure conditions. Our results show 
that an {\it ab initio} rigid water model yields faster 
diffusion and radial 
distribution functions which are less structured 
than those found with a flexible 
model. Overall, the properties computed
with the rigid model are in better agreement with 
experiment than those determined with a flexible model. 
Possible reasons for this improved agreement are 
discussed.  In addition, 
we present a localized orbital analysis of the trajectories 
obtained with both a rigid and flexible water model, and we 
demonstrate that the large dipole moment changes in going from 
the gas to the liquid phase are not significantly altered by 
the rigid water approximation. The use of a rigid water model 
in {\it ab initio} simulations opens the way to much longer simulations 
of solutions where chemical reactions and dissociation effects 
do not occur.

\section{Methods}

In order to examine how the structural and dynamical properties of water
are altered by a rigid water approximation, we have performed a series of 
first principle molecular dynamics simulations \cite{gp} of water with and without 
intramolecular bond and angle constraints under ambient and high pressure 
and temperature conditions. 
The simulations consist of 54 water molecules in a 
periodically repeated cubic cell with a lateral dimension of 
either 11.74 \AA\ or 10.10 \AA, which correspond to 
densities of 1.00 g/cc and 1.57 g/cc, respectively. At each density, 
we have compared simulations where the intra-molecular geometry of
the water molecules are rigid to those where the geometries of 
the water molecules are fully flexible. 

\begin{table}
\caption{Details of the simulations.}
%\vskip .1in
\begin{ruledtabular}
\begin{tabular}{clccccc}
Name & Geometry & $\rho$ (g/cc) & T (K) & dt (fs) & $\mu$ (a.u.) & time (ps) \\
\tableline
A & Rigid    & 0.997 & 326 & 0.07 &  765 & 16.8 \\
B & Rigid    & 0.997 & 315 & 0.24 & 1100 & 24.5 \\
C & Flexible & 0.997 & 291 & 0.07 &  340 & 19.8 \\
D & Rigid    & 1.570 & 603 & 0.07 &  765 & 18.6 \\
E & Flexible & 1.570 & 600 & 0.05 &  340 &  3.0 \\
\end{tabular}
\end{ruledtabular}
\label{details}
\end{table}

In each simulation, the electronic structure was described within
density functional theory (DFT) \cite{phohenberg64,wkohn65}
with the PBE generalized gradient approximation \cite{jperdew98}.
The valence wavefunctions and charge density were 
expanded in a plane wave basis, which was truncated in reciprocal 
space at 85 and 340 Ry, respectively. Norm-conserving pseudopotentials 
of the Hamman type were used to describe valence-core interactions 
\cite{dhamann89,lkleinman82}. 

The simulations were performed with the CP technique \cite{rcar85},
which is based on the use of a Lagrangian that couples
together the system's electronic and ionic degrees of freedom,
\begin{eqnarray}
L_{CP} & = &
\mu\sum_{i}f_i\int d{\bf r} \left|\dot{\Psi}_i\left({\bf r}\right)
\right|^2+\frac{1}{2}\sum_{I}M_I{\bf \dot{R}}_I^2-  \nonumber \\ & &
E_{KS}
\left[\left\{\Psi_i\right\},
{\bf R}_I\right]+ \nonumber \\ & &
\sum_{ij} \Lambda_{ij}\left(\int d{\bf r} \Psi_i^*\left({\bf r}\right)
\Psi_j\left({\bf r}\right)-\delta_{ij}\right).
\label{lagrangian}
\end{eqnarray}
In Eq.~\ref{lagrangian}, $\Psi_i\left(r\right)$ are the Kohn-Sham
orbitals, $\mu$ is a fictitious mass parameter used to evolve
the electronic degrees of freedom in time, $M_I$ are ion masses,
$E_{KS}$ is the Kohn-Sham energy, $f_i$ are occupation numbers, 
and $\Lambda_{ij}$ are Lagrange
multipliers, used to impose the orthonormality constraint
$\int\Psi_i^*\Psi_j=\delta_{ij}$.  The equations of motion derived
from Eq.~\ref{lagrangian} are,
\begin{equation}
\mu\ddot{\Psi}_i\left({\bf r},t\right)  =  -
\frac{\delta E}{\delta\Psi_i^*\left({\bf r},t\right)}
+\sum_j\Lambda_{ij}\Psi_j\left({\bf r},t\right) \label{eome} 
\end{equation}
and
\begin{equation}
M_I{\bf \ddot{R}}_I =  -\frac{\partial E}{\partial {\bf R}_I\left(t\right)}.
\label{eomi}
\end{equation}

Central to the CP method is the introduction of the fictitious
mass parameter, $\mu$, which enables the electronic and ionic trajectories
to be propagated simultaneously at each time step. The particular value
of $\mu$ is chosen so that the dynamics of the electronic degrees of 
freedom occurs on
a time scale that is much faster than, and thus decoupled from, the ion
dynamics. 

For a given system, an appropriate value of $\mu$ can be determined 
by approximating the dynamics of the orbitals generated by 
Eq.~\ref{eome} as a superposition of oscillators with frequencies,
\cite{gpastore91,ggalli91}
\begin{equation}
\omega_{ij}=\left[\frac{2\left(\epsilon_{j}-\epsilon_{i}\right)}
{\mu}\right]^{1/2},
\label{eq_elec}
\end{equation}
where $\epsilon_i$ and $\epsilon_j$ are Kohn-Sham eigenvalues of 
occupied and unoccupied states, respectively. The lowest frequency 
obtained from Eq.~\ref{eq_elec} occurs when $\epsilon_i$ is the 
highest occupied state (HOMO) and $\epsilon_j$ is the lowest unoccupied 
state (LUMO). As discussed in Ref.~\onlinecite{gpastore91}, although 
Eq.~\ref{eq_elec} is a rather crude approximation of the true orbital 
dynamics, it still provides a useful estimate for selecting an 
appropriate value of $\mu$ \cite{jgrossman03}. 

In the case of water, the highest ionic frequency is due to O-H stretching modes,
which are at $\sim$3500 cm$^{-1}$ within density functional theory. 
In addition, the HOMO-LUMO gap for water 
within density functional theory is approximately 
4.6 eV \cite{klaasonen93}.
As discussed in Ref.~\onlinecite{jgrossman03}, this means that 
values of $\mu\sim$340 au are needed to ensure a 
clear separation between the ionic and electronic degrees of freedom in 
CP simulations of flexible, protonated water. However, when the rigid water approximation 
is used, much larger values of $\mu$ can be used because the bond 
distance and angle constraints suppress the high frequency ionic motions. 
The various values of $\mu$ that we have used for both the rigid and 
the flexible water simulations are listed in Table \ref{details}.

The initial starting configurations of the simulations were generated 
from previous simulations of flexible water \cite{eschwegler01} by 
scaling the intramolecular O-H distances and H-O-H angles,
while leaving the spatial orientations of the molecules unchanged. 
The particular values for the geometry, $0.9926$ \AA\ for the O-H 
distance and $104.6^\circ$ for the H-O-H 
angle, were chosen to be the same as the peak values for the 
intra-molecular O-H distance and H-O-H angular distributions obtained 
in the previous flexible water simulations \cite{eschwegler01}.
These bond distances and angles are  similar to the 
ones used in the well-known TIP classical water models ($0.9572$ \AA,
$104.52^\circ$) \cite{mmahoney00}. In order to maintain the distance and angle 
constraints at every iteration, the SHAKE algorithm 
was used \cite{jryckaert77,mpallen87}.

In simulations A and B (Table \ref{details}), the samples were 
initially heated to a 
temperature of 450 K and cooled to 300 K for a period of
2 ps. In simulation D, the system was heated to a temperature of 
900 K and cooled to 600 K over 4 ps. The thermostats were then 
removed and the simulations were run under constant energy 
conditions for the times listed in Table \ref{details}.

Maximally localized Wannier functions (MLWF)
\cite{nmarzari97,psilvestrelli98}, analogous to the 
orbitals obtained by the Boys localization procedure
\cite{sboys60}, were used to 
determine how the rigid water approximation may influence the 
electronic structure of the water 
molecules. By using a recently proposed joint approximate diagonalization 
scheme, we were able to compute the MLWFs ``on-the-fly" for a large number of 
configurations \cite{fgygi03}. Following the procedure proposed 
by Silvestreli {\it et al.}~\cite{psilvestrelli99}, the centers of the 
MLFWs have been used to compute an approximate dipole moment of 
each of the water molecules in the simulation. 

\section{Results}

\begin{figure}
\centerline{
\rotatebox{-90}{\resizebox{2.9in}{!}{\includegraphics{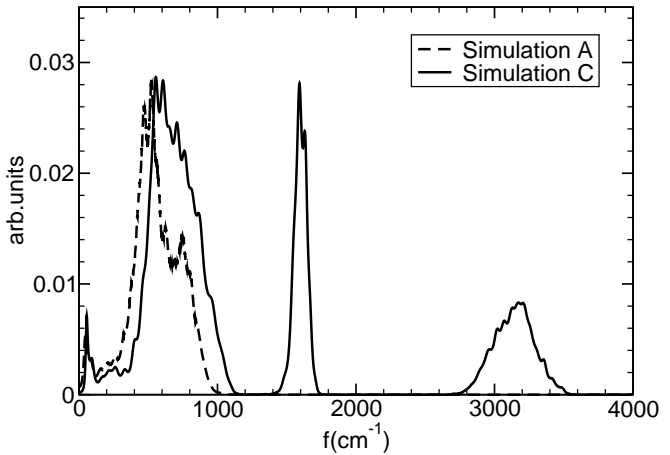}}}
}
\caption{The computed power spectrum of rigid and flexible liquid water
(see text). The 
dashed line corresponds to the rigid simulation A and the solid line to 
the flexible simulation C.}
\label{powerspecALL}
\end{figure}

The primary effect of the rigid water approximation is best 
illustrated by comparing the vibrational spectra obtained from a
rigid and a flexible water simulation. 
In Fig.~\ref{powerspecALL}, the power spectra collected from 
simulations A and C are shown. The spectra were computed 
directly from the velocity autocorrelation function 
with the maximum entropy method. The two highest frequency modes 
in the flexible water simulation C are completely absent in simulation A. 
These modes at 1600 cm$^{-1}$ and 3150 cm$^{-1}$ correspond to 
intra-molecular H-O-H bending and O-H stretching, respectively. By 
suppressing the high frequency modes large 
values of $\mu$ can be used, which in a fully flexible simulation 
would normally lead to a coupling of the electronic and ionic degrees of 
freedom, and in turn to a breakdown in the adiabiticity of the 
CP dynamics \cite{jgrossman03}.

In addition to the obvious absence of high 
frequency modes in the rigid water simulation, there are small 
differences in the lower frequency range of the spectra below 
$\sim$1000 cm$^{-1}$. 
For example, the librational modes are slightly shifted 
to lower frequencies by approximately 60 wavenumbers in the 
rigid water simulation A. Overall, the differences in the low 
frequency region of the rigid and flexible water simulations are 
small. 

\begin{figure}
\centerline{
\rotatebox{-90}{\resizebox{2.9in}{!}{\includegraphics{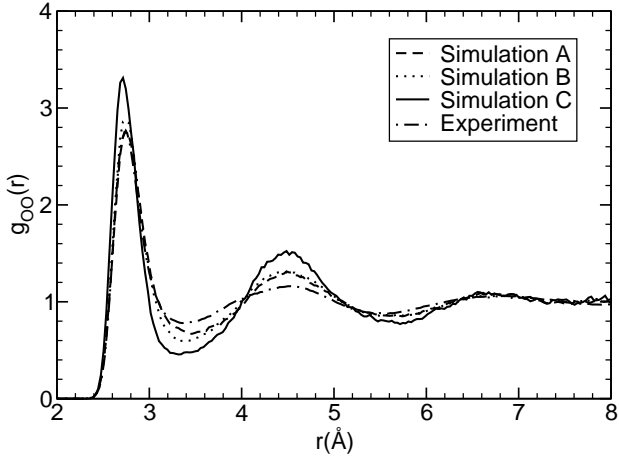}}}
}
\caption{The oxygen-oxygen radial distribution function of water at
ambient conditions. The dashed and dotted lines correspond to the
rigid water simulations A and B, respectively, the solid line to the
flexible water simulation C, and the dot-dashed line to
experiment \cite{asoper00b}.}
\label{gOO}
\end{figure}

In Fig.~\ref{gOO}, the oxygen-oxygen radial distribution functions 
(RDFs) obtained from simulations A, B and C are shown. Also displayed 
in Fig.~\ref{gOO}, is the latest experimental oxygen-oxygen RDF obtained 
by an analysis 
of neutron diffraction data \cite{asoper00b}, which is nearly identical 
to the RDF determined from a recent x-ray diffraction study of 
water \cite{jsorenson00}.  The small 
differences between simulations A and B indicate that 
the choice of $dt$=0.07 or 0.24 fs has 
little effect on the oxygen-oxygen RDF when the rigid water 
approximation is used.  The differences 
in the height of the first peak and the first minimum in the RDFs 
are most likely due to the $\sim$10 K higher average temperature of 
simulation A over simulation B (see Table \ref{details}). However, 
as discussed in Ref.~\onlinecite{jgrossman03}, for the simulation times used 
here (15-25 ps), differences of this magnitude are within 
the expected error bars. In Fig.~\ref{gOO} we also show the 
oxygen-oxygen RDF obtained from simulation C with flexible water 
molecules. The differences between the rigid and the flexible water simulations 
are significant. The first peak height is approximately 
0.4 higher in simulation C than in simulations A and B, and is 
shifted inward by 0.03 \AA. In general, the RDF peak heights and 
depths are decreased in the rigid water simulations,
showing that the rigid water approximation causes an 
overall decrease in the structure of the liquid. The same 
trend in terms of the peak heights and depths has been 
observed in flexible and rigid water simulations by using the 
SPC classical interaction potential \cite{jlobaugh97}. 

\begin{figure}
\centerline{
\rotatebox{-90}{\resizebox{2.9in}{!}{\includegraphics{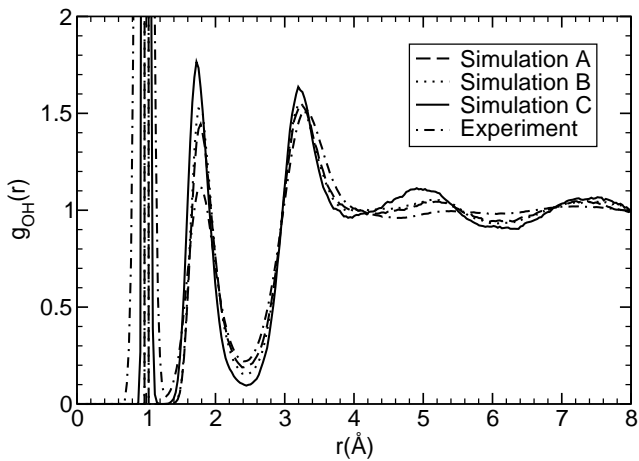}}}
}
\caption{The oxygen-hydrogen radial distribution function of water at
ambient conditions. The dashed and dotted lines correspond to the
rigid water simulations A and B, respectively, the solid line to
the flexible water simulation C, and the dot-dashed line to
experiment \cite{asoper00b}.}
\label{gOH}
\end{figure}

\begin{figure}
\centerline{
\rotatebox{-90}{\resizebox{2.9in}{!}{\includegraphics{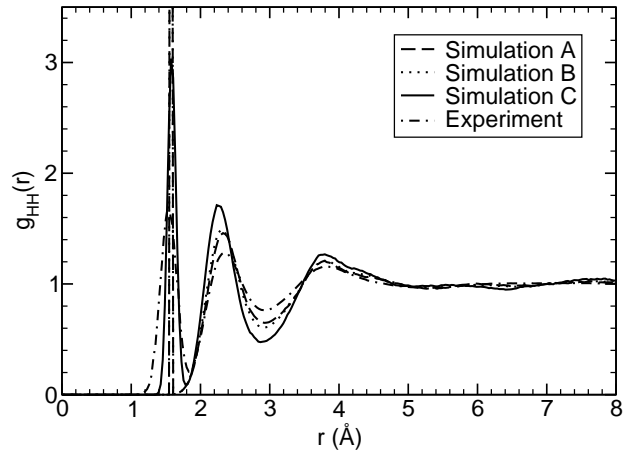}}}
}
\caption{The hydrogen-hydrogen radial distribution function of water at
ambient conditions. The dashed and dotted lines correspond to the
rigid water simulations A and B, respectively, the solid line to
the flexible water simulation C, and the dot-dashed line to
experiment \cite{asoper00b}.}
\label{gHH}
\end{figure}

The tendency of the rigid water approximation to decrease the 
structure of water can also be seen in the oxygen-hydrogen and the 
hydrogen-hydrogen RDFs, as shown in Figs.~\ref{gOH} and \ref{gHH}. 
It is interesting to note that in simulation C the second peak in the 
oxygen-hydrogen RDF at r$\sim$2.8 \AA\ is larger in height than the 
third peak at r$\sim$3.2 \AA, whereas in simulations A and B, as well as 
in the experimental measurement \cite{asoper00b}, the 
second peak is smaller than the corresponding third peak. Since the 
second oxygen-hydrogen peak coincides with pairs of hydrogen
bonded water molecules in the liquid, the relative heights of the 
second and third peaks have been used as an indicator of 
the amount of hydrogen bonding in the liquid \cite{jlobaugh97}.

Overall, Figs.~\ref{gOO} to \ref{gHH} suggest that the main effect of 
the rigid water approximation is to remove a large fraction of the 
over-structure that is characteristic of flexible water as described within 
DFT both by CP \cite{jgrossman03} and BO molecular 
dynamics \cite{dasthagiri03}, and in general, the rigid
water distribution functions appear to be in better agreement with experiment. 
There are several possible reasons for this improved agreement.
For example, it is possible that the rigid water approximation 
mimics some of the effects coming from the quantum nature of protons,
which are explicitly ignored in simulations that allow for intramolecular 
flexibility with classical dynamics \cite{itironi96,jlobaugh97,bhess02}.
More specifically, in Ref.~\onlinecite{itironi96} it was pointed out that at a 
temperature of 
300 K, $k_BT\sim200$ cm$^{-1}$, whereas the high frequency intramolecular
modes in water range from $\sim$1000 to 3500 cm$^{-1}$. In 
the quantum system the amount of thermal energy available to excite 
vibrational modes is much smaller than the lowest possible intramolecular 
vibrational excitations, 
\begin{equation}
\hbar\omega \gg k_BT.
\end{equation}
This indicates that the real quantum system will essentially be 
restricted to its vibrational ground state at 300 K. Therefore, the 
quantum system may be better described by a classical rigid water model 
than by a classical
flexible model, despite the fact that in the quantum system the protons 
become delocalized compared to the classical case.

The effect of the proton quantum motion can be considered with
path-integral (PI) methods \cite{rfeynman65}. To date, 
all PI simulations with empirical interaction potentials 
have found an overall softening in the RDFs of water at 
ambient conditions
\cite{rkuharski85,gbuono91,jlobaugh97,hstern01,mmahoney01}. 
As pointed out in Ref. \onlinecite{gbuono91} these structural 
changes are quite similar to a 50 K increase in the 
simulation temperature.  
However, it is not clear if the lack of quantum effects can 
fully account for all of the overstructure that appears to 
be characteristic in the DFT/GGA treatment of water. 
In addition, a recent PI-DFT simulation \cite{bchen03} has found 
the surprising result that quantum effects may enhance hydrogen 
bonding in water.

In order to examine how the dynamical properties of water are 
affected by the rigid water approximation, we have estimated
the diffusion coefficients, D, for the simulations of rigid and flexible 
water. To compute the self diffusion coefficient, the mean square 
displacement (MSD) of the oxygen atoms in the simulations were tracked 
as a function of the simulation time. The statistical sampling of the
simulation data was improved by defining multiple starting configurations
separated by 4 fs. The slope of the resulting MSD 
in the range of 1 to 10 ps was then determined and 
used to estimate D according to the Einstein relation \cite{mpallen87},
\begin{equation}
6D_1 = \lim_{t\rightarrow\infty}\frac{d}{dt}
\left<\left|{\bf r}_i\left(t\right)
-{\bf r}_i\left(0\right)\right|^2\right>.
\label{diffcoef}
\end{equation}

The diffusion coefficient can also be estimated 
via integration of the oxygen velocity autocorrelation function as
\begin{equation}
D_2 =\frac{1}{3}\int_0^\infty dt \left<{\bf v}\left(t\right)
\cdot {\bf v}\left(0\right)\right>,
\label{vint}
\end{equation}
which for the simulation times used here yields nearly identical results 
as Eq.~\ref{diffcoef}. The measured diffusion coefficients for the 
different simulations as computed by Eqs.~\ref{diffcoef} and \ref{vint}
are reported in Table \ref{diffusion}. For the
rigid water simulations A and B, the diffusion coefficients are
approximately 6.2 and 3.2 times larger, respectively, than the flexible 
water simulation C. 
Although simulations A and B were both performed at higher temperatures
than simulation C, the observed increases in the diffusion coefficients 
are outside the increase (1.4 to 1.7 fold) expected for a 15 to 
26$^\circ$ increase in temperature over ambient, based on experimental 
data \cite{kkrynicki78}. Given that the rigid water approximation results in 
a less structured liquid than the flexible one, 
it is not surprising that it also leads to faster diffusion, which is in closer 
agreement with the experimental measurement of $2.4\times10^{-5}cm^{2}/s$
\cite{rmills73,kkrynicki78}.

\begin{table}
\caption{Summary of computed diffusion coefficients. D$_1$ corresponds
to the diffusion coefficient as estimated by the Einstein relation and
D$_2$ is from integrating the velocity autocorrelation function. }
\vskip .1in
\begin{ruledtabular}
\begin{tabular}{ccc}
Simulation &
D$_1$ (10$^{-5}$cm$^2$/s) &
D$_2$ (10$^{-5}$cm$^2$/s) \\
\tableline
A  &   2.5  & 2.4 \\
B  &   1.3  & 1.2 \\
C  &   0.3  & 0.4 \\
D  &   2.5  & 2.5 
\end{tabular}
\end{ruledtabular}
\label{diffusion}
\end{table}

\begin{figure}
\centerline{
\rotatebox{-90}{\resizebox{2.9in}{!}{\includegraphics{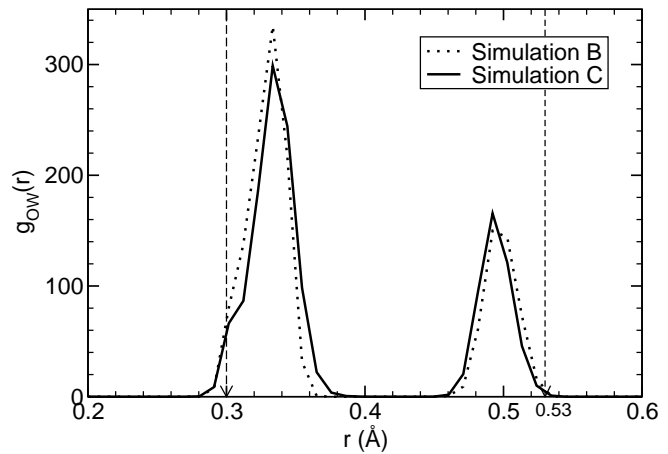}}}
}
\caption{The oxygen-MLWF center radial distribution function for rigid 
and flexible water. The dotted line corresponds to the rigid simulation B and the 
solid line to the flexible simulation C. The dashed vertical lines represent the 
location of MLWFs for an isolated water molecule.}
\label{gOW}
\end{figure}

In addition to possible changes in structure and dynamics, 
we have examined how the rigid molecule approximation changes the 
electronic properties of water. In order to do this 
we have performed a localized orbital analysis by computing the 
MLWFs for a series of well-separated snapshots from the rigid 
water simulation B and the flexible water simulation C.
Within the pseudopotential approximation, 
there are four doubly occupied MLWFs around each of the water 
molecules in the simulations. Two of the MLWFs are localized on the 
oxygen-hydrogen covalent bonds, and the other two are localized 
on the lone-pair locations of the oxygen atoms. Given the large 
amount of data that the MLWFs represent, in the following, we only 
consider the centers of the MLWFs rather than the orbitals 
themselves. In Fig.~\ref{gOW}, the oxygen-MLWF center RDFs for 
simulations B and C are shown. The RDFs consist of two 
distinct distributions centered at r$\sim$0.33 \AA\ and 
r$\sim$0.49 \AA, which correspond to lone-pair and 
covalent bond locations, respectively. For comparison, the dashed
vertical lines in Fig.~\ref{gOW} represent the locations of the 
MLWF centers around the oxygen atom of an isolated gas phase water 
molecule. Surprisingly, the rigid water approximation does not 
significantly alter the large changes in the MLWF centers that are
expected when going from an isolated water molecule to the liquid 
state. As can be seen in Fig.~\ref{gOW}, the lone pair distributions
are shifted away from the oxygen atoms by $\sim$0.03 \AA, and the 
covalent bond distributions are shifted toward the oxygen atoms 
by $\sim$0.04 \AA. 

As proposed by Silvestrelli {\it et al.} \cite{psilvestrelli99},
an approximate dipole moment for each water molecule in the liquid 
can be defined by assigning the total charge of each MLWF to a point 
charge located at its corresponding center. Because the MLWF on 
neighboring water molecules in the liquid do not significantly 
overlap, this provides a less ambiguous definition of the 
molecular dipole moments than arbitrarily assigning electron density to 
individual water molecules. As pointed out in Ref.~\onlinecite{apasquarello03}, 
dipole moments computed in this manner from static configurations 
may not be representative of the experimentally measured dipole moments 
in the fluid. However, the 
MLWF dipole moments are useful for examining qualitative 
differences in the polarization of water as a function of different 
approximations or of solutes present in the liquid. 

\begin{figure}
\centerline{
\rotatebox{-90}{\resizebox{2.9in}{!}{\includegraphics{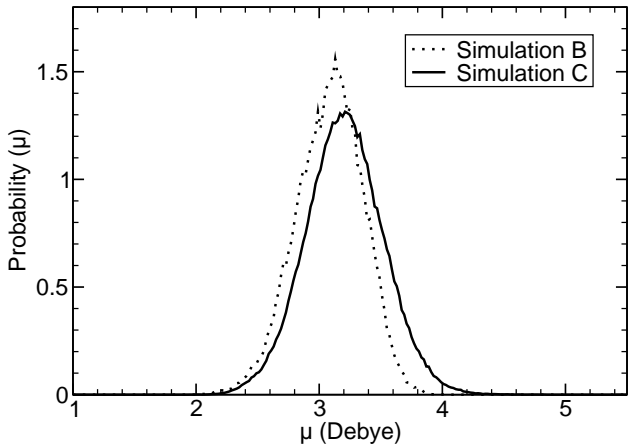}}}
}
\caption{The probability distribution of molecular dipole moments of water
computed from MLWF centers. The dotted line 
corresponds to the rigid simulation B and 
the solid line to the flexible simulation C.}
\label{dipole}
\end{figure}

In Fig.~\ref{dipole}, the probability distributions of molecular dipole 
moments for rigid and flexible water systems calculated from the MLWF
centers are shown. Fig.~\ref{dipole} indicates that the 
rigid water approximation has a rather small effect on distribution of 
dipole moments in the liquid. In particular, the average moment in 
simulation B is shifted to 3.08 Debye as compared to 3.20 Debye in
simulation C. Apparently, an explicit description of high frequency 
O-H stretch and H-O-H 
bending modes is not necessary to reproduce the broad range 
of moments that are characteristic of the liquid. It is also 
interesting to note that the latest experimental estimate based on 
an analysis of the x-ray structure factor of water indicates that the 
dipole moment of water in the liquid is 2.9 Debye \cite{ybadyal00}, which is 
in closer agreement with the rigid water model than the flexible simulation. 

The decrease in the average dipole moment obtained from the 
rigid water approximation offers another explanation for 
the observed softening of the liquid structure. In addition to 
mimicking quantum effects, it is possible that
the rigid water approximation to some extent corrects for the 
general tendency of simple GGA-based functionals to overestimate 
the polarizability of molecules 
\cite{smcdowell95,pcalaminici98,acohen99,ccaillie00}. 
For example, the static isotropic polarizability of an 
isolated water molecule is 10.74 au with the PBE functional as 
compared to the experimental value of 9.64 au \cite{ccaillie00}. 
It is interesting to note that hybrid DFT functionals,
which include some amount of Hartree-Fock exchange, appear to significantly 
improve on the polarizability of water. In particular, the average 
polarizability of the water molecule is 9.78 au with hybrid PBE0 
functional \cite{ccaillie00}. 

\begin{figure}
\centerline{
\rotatebox{-90}{\resizebox{2.9in}{!}{\includegraphics{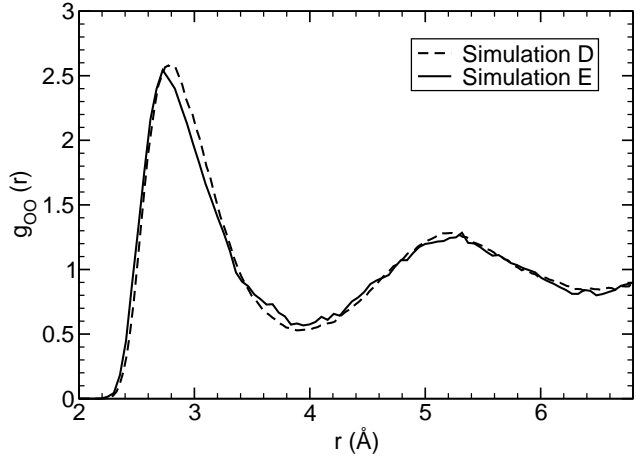}}}
}
\caption{The oxygen-oxygen radial distribution function of water at
$\rho$=1.57 g/cc. The dashed line is from the rigid water simulation D,
and the solid line is from the flexible water simulation E.}
\label{gOO-HP}
\end{figure}

\begin{figure}
\centerline{
\rotatebox{-90}{\resizebox{2.9in}{!}{\includegraphics{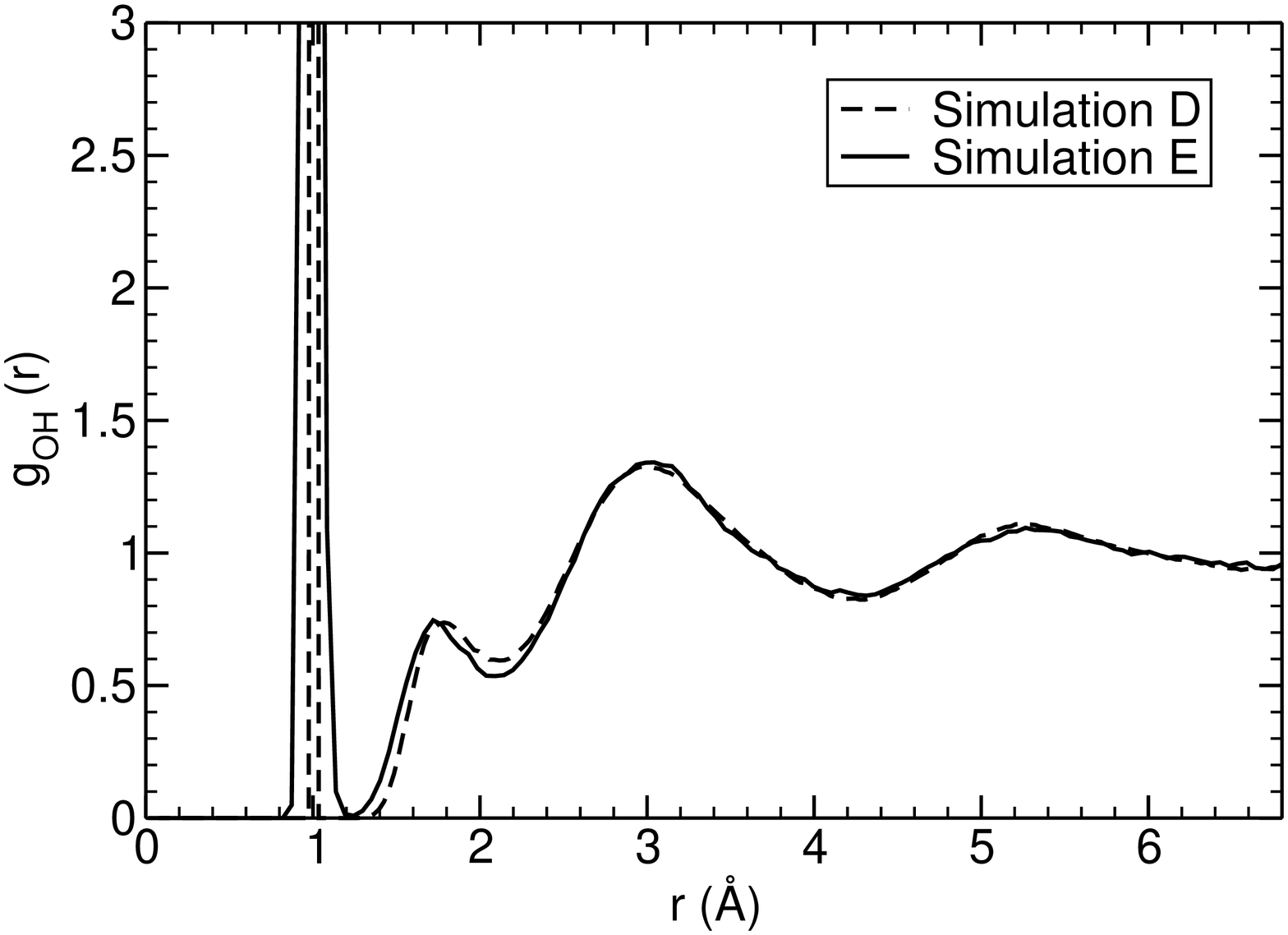}}}
}
\caption{The oxygen-hydrogen radial distribution function of water at
$\rho$=1.57 g/cc. The dashed line is from the rigid water simulation D,
and the solid line is from the flexible water simulation E.}
\label{gOH-HP}
\end{figure}

\begin{figure}
\centerline{
\rotatebox{-90}{\resizebox{2.9in}{!}{\includegraphics{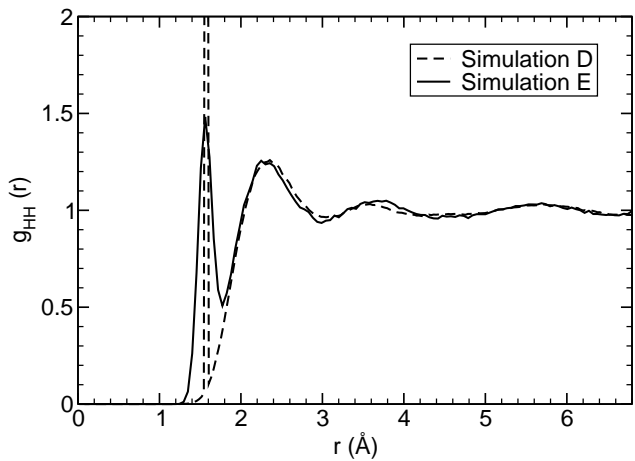}}}
}
\caption{The hydrogen-hydrogen radial distribution function of water at
$\rho$=1.57 g/cc. The dashed line is from the rigid water simulation D,
and the solid line is from the flexible water simulation E.}
\label{gHH-HP}
\end{figure}

In addition to water under ambient conditions, we have also examined how 
the rigid water approximation affects the properties of water under 
extreme temperatures and pressures. In particular, we have performed 
a simulation of rigid water (simulation D) and of flexible water (simulation 
E) at a density of 1.57 g/cc and 
an average temperature of $\sim$600 K. These high density and temperature 
conditions correspond to a regime where molecular dissociation is still 
considered a rare 
event \cite{eschwegler00,eschwegler01}. However, the pressure ($\sim$10 GPa) 
is high enough to cause a large increase in the nearest neighbor coordination 
of each water from 4.5 at ambient conditions to nearly 13 at high pressure
\cite{eschwegler00}.
The oxygen-oxygen, oxygen-hydrogen and hydrogen-hydrogen RDFs for 
simulations D and E are compared in Figs.~\ref{gOO-HP} to \ref{gHH-HP}.
Except for the expected intramolecular differences due to the constraints, 
the RDFs obtained from simulations D and E are remarkably similar. In 
particular, both the large increase in the number of nearest neighbors as
well as the stiffness of the first peak in g$_{\rm OO}$(r) as a function of
compression are reproduced by the rigid water model \cite{eschwegler00}. 
As higher densities and temperatures are considered, intramolecular 
dissociation will become an common event
in flexible water simulations \cite{eschwegler01} and the rigid 
water approximation is expected to be inappropriate for the 
description of the liquid.

\section{Conclusion}

In summary, we have used a series of Car-Parrinello molecular dynamics 
simulations to examine how the rigid water approximation affects the 
computed
properties of water in the liquid state at ambient conditions. 
In agreement with 
previous observations based on empirical interaction potentials
\cite{jlobaugh97}, 
the rigid water approximation is found to cause an overall
decrease in structure and an increase in diffusion of the liquid. 
These changes result in properties that are in better 
agreement with experimental measurements than the corresponding 
first principles simulations with flexible water molecules. 
At higher temperatures 
and densities in a regime where intramolecular dissociation is still
a rare event, the differences between simulations where 
water molecules are either rigid or flexible become negligible. 

In addition to an improved structural and dynamical description of water, 
the rigid water model enables the use of time steps as large as 
0.24 fs within the Car-Parrinello scheme (i.e. $\sim$3 times larger than 
in a flexible water simulation). A similar conclusion was reached in 
Ref.~\onlinecite{etsuchida01} for first-principle simulations of a 
cytosine molecule in the gas phase. This represents an important
advantage for first-principle simulations of aqueous solutions where chemical
reactions do not occur, and
opens up the possibility of investigating phenomena that take place on a 
long timescale. For example, understanding how water orients around a 
hydrophobic solute may require simulations of the order of 100 to 200 ps.
The rigid water approximation presented here may prove to be an accurate and 
efficient approach for describing the interaction between a hydrophobic solute 
and water within a first-principles context. \\

The authors would like to thank J.~C.~Grossman for many useful
discussions.  This work was performed under the auspices of 
the U.S.~Dept.~of Energy at the University of California/Lawrence 
Livermore National Laboratory under contract no.~W-7405-Eng-48.

\end{document}